\documentclass[aps,prfluids,onecolumn,longbibliography]{revtex4-2}

\usepackage{graphicx,color,amsmath,amssymb,hyperref}
\def\dd{{\, \rm{d}}}
\def\dr{{\rm{d}}}

\def\p{\partial}

\def\beq{\begin{equation}}
\def\eeq{\end{equation}}
\def\la{\label}
\def\ii{{\rm i}}

\def\r#1{(\ref{#1})}
\newcommand{\mylab}[3]{\raisebox{#2}[0mm][0mm]{%
\makebox[0mm][l]{\hspace*{#1}{#3}}}}%
\def\utau{u_\tau}
\def\retau{Re_\tau}

\def\hkappa{\widehat{\kappa}}
\def\hu{\widehat{q}}

\def\figpath{./}
\def\spacce#1{\hskip #1pt}
\def\drawline#1#2{\raise 2.5pt\vbox{\hrule width #1pt height #2pt}}
\def\solid{\drawline{24}{.5}\nobreak}

\def\bdash{\hbox{\drawline{5.8}{.5}\spacce{2}}}

\def\dashed{\bdash\bdash\bdash\nobreak}

\def\trian{\raise 1.25pt\hbox{$\scriptstyle\triangle$}\nobreak}

\def\dtrian{\raise 1.25pt\hbox%
{$\scriptscriptstyle\bigtriangledown$}\nobreak}

\def\squar{\raise 1.25pt\hbox{$\scriptstyle\Box$}\nobreak}

\def\diamon{\raise 1.25pt\hbox{$\scriptstyle\diamond$}\nobreak}

\newcommand{\soliddtrian}{$\blacktriangledown$\nobreak}

\def\linedtri1{\hbox{\bdash\hspace{-1.6mm}$\bigtriangleup$\hspace{-0.8mm}\bdash}\nobreak}
\def\soliddtrian1{$\blacktriangledown$\nobreak}
\def\solidrtrian2{$\blacktriangleright$\nobreak}
\def\solidltrian3{$\blacktriangleleft$\nobreak}

\begin{document}

\title{Chaos, coherence and turbulence}

\author{Javier Jim\'enez}
\email[]{javier.jimenezs@upm.es}
\affiliation{Aeronautics, Universidad Polit\'ecnica de Madrid, 28040 Madrid, Spain }

\date{\today}

\begin{abstract}
This paper is a personal overview of the efforts over the last half century to understand fluid
turbulence in terms of simpler coherent units. The consequences of chaos and the concept of
coherence are first reviewed, using examples from free-shear and wall-bounded shear flows, and
including how the simplifications due to coherent structures have been useful in the
conceptualization and control of turbulence. It is remarked that, even if this approach has
revolutionized our understanding of the flow, most of turbulence cannot yet be described by
structures. This includes cascades, both direct and inverse, and possibly junk turbulence, whose
role, if any, is currently unknown. This part of the paper is mostly a catalog of questions, some of
them answered and others still open. A second part of the paper examines which new techniques can be
expected to help in attacking the open questions, and which, in the opinion of the author, are the
strengths and limitations of current approaches, such as data-driven science and causal inference.
\end{abstract}


\maketitle

\section{Introduction}\la{sec:hist}

This article grew out of the Otto Laporte lecture delivered by the author at the 2024 meeting of the
Division of Fluid Dynamics of the American Physical Society, and reflects a personal view of what
the recent history of turbulence research has been, and of what I think remains to be done. As such,
I identify particularly personal opinions by occasionally taking the liberty of using the first
person singular, and readers may want to supplement those parts of the paper with their own points
of view. Parts of this article could probably have been written about other subjects, because
turbulence has not been immune to the changes in science and engineering in the past century, nor is
it isolated from the changes taking place today. However, turbulence is the only subject in which I
consider myself barely proficient enough to have an opinion and, as I will hopefully show below, it
is also important enough by itself and in its relations to other fields to be scientifically
interesting and technologically relevant.
 
Turbulence typically appears when the Reynolds number exceeds $O(100-1000)$, which is often
the case in natural and industrial flows because the kinematic viscosity of air is $10^{-5}$, and
that of water is $10^{-6}$, when expressed in the units in which we function (meters and seconds).
Turbulence is not very relevant to the life of bacteria, but most macroscopic industrial or natural
flows tend to be turbulent, and this modifies their behavior enough to consider them a different
flow regime. In some cases, this is undesirable, creating friction and dissipating energy in
vehicles and pipes, or increasing noise in jets. In other occasions, its effect is desirable or even
indispensable, as when it dissipates energy in parachutes, spillways or brakes; landing a commercial
plane involves getting rid of the potential energy of a hundred tons falling across several thousand
meters. Most mixing in nature, jets, chemical reactors and coffee cups is done by turbulence,
including the relatively uniform climate of our planet. The temperature difference between the poles
and the equator would be unbearable without the turbulent transport of heat and water vapor, and the
well-mixed troposphere in which we breathe and live would not exist without convective turbulence.

There are excellent textbooks dealing with what is known about turbulence
\cite{bat53,tenn,hinze,les97,pope:00}. They will not be reproduced here, and the reader is
encouraged to consult them for the fundamentals of the field.
 
The dynamics of turbulence can be roughly divided into three ranges. Energy enters the fluctuations
from a large-scale reservoir, typically shear or a density gradient. The incoming energy cannot be
dissipated at these scales, which become intense and distribute their energy to larger or smaller
scales. This distribution mechanism, often described as a cascade, characterizes the second range of
scales, and may have several subranges. The third range is where energy dissipates or, in
out-of-equilibrium or very inhomogeneous situations, accumulates or spreads. These are typically the
smallest scales in the flow although, in cases in which the cascade is reversed, dissipation at
large scales is also required. Although cascades are the most characteristic feature of turbulence,
this paper deals mainly with the energy input range, which is non-universal and therefore full of
open questions. We do this by reviewing in \S\ref{sec:FSF} and \S\ref{sec:orr} the
large-scale dynamics of free-shear and wall-bounded flows, respectively, both of which draw their
energy from the shear although, as we will see, in different ways. We will find that the
energy-input structures are coherent in the sense that they are relatively independent of the rest
of the flow, including of the cascade. The latter and the incoherent part of turbulence are relegated
to \S\ref{sec:incoherent} and, as a consequence, the paper may look like a set of open questions,
rather than as a collection of answers. But, in spite of its long history, open questions 
abound in turbulence, and the author hopes that enough readers agree with him that trying to predict
the future is more fun (although more risky) than recalling the past.

\subsection{Some history}\la{sec:history}

Humanity has probably been aware of turbulence from the very beginning. Hominins must have known
that swimming across a river is difficult because of its violent eddies, sailors were certainly
aware of the unsteadiness of the wind, and ancient civilizations built impressive waterworks that
betray an empirical knowledge of the flow behavior, but the modern study of turbulence began with the
industrial revolution in the mid nineteenth century. At that time, only integral values like the friction
coefficient or the mass flux were important, and most early results are empirical observations in
the form of tables and plots \cite{hag39,dar55}. However, it did not take long before some form
of theory became necessary. The pressure loss in the circulation of blood in capillaries had been
clarified in the 1840s \cite{pois46}, including the central role of the viscosity coefficient, and
the Navier--Stokes equations had been written at about the same time \cite{Navier22,Stokes45},
building on the earlier work of \citet{euler:1757}. It came as a surprise that larger pipes, as
for example those in municipal water distribution networks \cite{hag39,dar55}, did not follow the
same laws as capillaries. Even more disturbing was that their friction coefficient, what we would
today call the turbulence dissipation rate, turned out to be independent of the viscosity
coefficient that had been shown to be so important in smaller vessels \cite{hag54}. It was soon
hypothesized that the velocity fluctuations in turbulent flows were connected with the unexplained
dissipation \cite{bous77}, and this led to the separation of mean and fluctuating components
\cite{rey94}, and to the first models of an eddy viscosity \cite{bous97}.

The next fifty years tended to view turbulence as the `laminar-like' flow of a strange fluid, whose
properties are determined by the statistics of the fluctuations, which were themselves seen as
largely stochastic \cite{bat53,tenn,hinze,les97}. A whole branch of turbulence research continues
this tradition to this day \cite{pope:00}, and is responsible for much of what we know as
the industrially relevant Reynolds-averaged Navier--Stokes (RANS) models \cite{launder:75}.

But estimating the behavior of a turbulent fluid is much simpler than understanding how this
behavior comes about. Conceptual models for how dissipation becomes independent of the dissipative
terms of the Navier--Stokes equations had to wait for the cascade models of the early twentieth
century \cite{rich22,kol41}, themselves largely statistical, and for the explanation of the relevant
limiting process by \citet{onsag49}. However, the approximations required to make airplanes fly or
to estimate how much pressure needs to be provided in water and gas distribution networks are
relatively independent of the detailed flow physics and, although the technology keeps improving,
the engineering description of turbulence using RANS was essentially complete half a century ago
\cite{launder:75}. This being the case, one could question why should we continue to worry about
turbulence. The following are some reasons important to the author. Readers should probably ask
themselves how would they rewrite the following lines.
\begin{itemize}
\item We may just be curious about it. This places turbulence research in the realm of fundamental
science, or even of `poetry'. Both are long-term investments and, although some variant of this is
probably the primary motivation for many of us, it raises the practical question of whether
turbulence should also be considered an important subject in engineering.

\item We may aim at better design, or maybe at control. Concrete and steel are mature technologies,
but building a really tall skyscraper requires specialty materials. In the same way, controlling, rather than
describing, turbulence requires physics, and so does, for example, delaying or avoiding transition.

\item Turbulence can help us with other physical problems, because it is one of the few examples of
a high-dimensional chaotic dynamical system for which we have detailed observations, simulations,
and experiments over which we have some measure of control.
\end{itemize}

In principle, even the more fundamental questions in this list should not pose an insurmountable
problem, because we believe that we know the equations of motion of a fluid, and their `direct'
numerical simulation (DNS) is well under control. However, DNS is too expensive for everyday use, and
it can be argued that it does not solve the problem of understanding. A typical modern DNS of a
canonical flow at a reasonable Reynolds number needs $O(10^{10})$ grid points, and must be run for
several thousands of time steps. Its output is thus $O(10^{13})$ numbers that have to be processed
before they are useful. Direct simulations are `black boxes' that reproduce the flow exactly, and
which, as a consequence, are as difficult to interpret as the flow itself. To compensate, and unlike
experiments, the results of a DNS are perfectly observable and complete, and can be stored and
processed as much as desired.

In the following pages we will center on the efforts during the past fifty years to reduce what we
have learned in this way to a more manageable representation that might be characterized as
understanding and, in the process, I will try to convince the reader that understanding is useful
from the point of view of engineering. As mentioned above, \S \ref{sec:coherent} summarizes
the evolution of the concept of coherence for free-shear flows, such as jets and mixing layers, and
for wall-bounded ones, such as boundary layers, channels and pipes. The open questions that we find
in them will drive us to consider the less coherent parts of turbulence in \S\ref{sec:incoherent},
as well as the effect of new processing techniques and theoretical approaches in \S\ref{sec:future}.
Section \ref{sec:conc} concludes with some guesses about the future development of the field.

\section{Coherent structures}\la{sec:coherent}

The late 1960’s and early 1970’s marked a turning point in turbulence research, because new
experiments revealed that some parts of the flow cannot be described as random fluctuations. This
triggered a transition from RANS and statistics to a more structural view, soon accelerated by the
detailed information that started to be available from computer simulations
\cite{orspat72,siggia81,Rogallo81,KMM87}.

The flow of a viscous fluid can be described by a finite number of degrees of freedom, or at least
converges exponentially to the neighborhood of a finite-dimensional manifold \cite{ConFoiTem:85}. It
can therefore be seen as a dynamical system governed by the deterministic Navier--Stokes equations,
\beq
\dr X/\dr t= f(X), 
\la{eq:DS0}
\eeq
where $X$ is some description of the state space, such as the velocity in a sufficiently dense grid
of points. For a three-dimensional turbulent flow, an estimate of the number of relevant degrees of freedom
scales as $Re^{9/4}$ \cite{landaufm}, which is usually a very large number. One
may wish to represent turbulence on a lower-dimensional projection, but the projected system is
not deterministic any more, which is probably one reason why early theories of turbulence tended to
be statistical, with randomness modeling the ignored degrees of freedom \cite{voltaire}. However, it
may happen that the projection on some submanifold, $X^*$, is almost deterministic for a limited
time, during which it approximately behaves independently from the neglected variables,
\beq
\dr X^*/\dr t= F(X^*)+\ldots \mbox{(smaller terms)}. 
\la{eq:DS1}
\eeq
This, I will call a coherent structure, understood as a submanifold that stays approximately
invariant for a substantial amount of time. If such a manifold exists, and if its dimensionality is
lower enough than that of the original system, some part of the flow can be described in terms of a
reduced set of variables, more simply than if the existence of structures had not been recognized.
Note that coherence remains useful even if it only describes part of the flow. For example, a
tornado is clearly coherent, and its coherence is relevant for predictions, but a tornado in Kansas
says little about rain in Montana. Coherence only requires that the tornado approximately behaves as
independent from that distant rain.

Other examples are Taylor--Couette flow and Rayleigh--B\'enard convection, where an initial
instability transitions to states that can be described in terms of rolls or cells. Although the
initial bifurcated states can hardly be called turbulent, the more complex secondary bifurcations
that grow from them can be analyzed for a while as small perturbations of the positions and
intensities of the transitional structures. Formally, we speak of a projection on a central manifold
\cite{Guck:Holm}; physically, of a description of the flow in terms of fewer degrees of freedom.

A common feature of all these flows is the existence of stable bifurcated states, whose
dimensionality is much lower than that of the full flow, and towards which the flow tends after some
time. Under those conditions, the flow can be described, up to some level, by properties on the
attractor. However, most `attractors' found in nature are not stable (and cannot therefore strictly
be called attractors). The system approaches them for a while, only to be repelled once it gets near
the central manifold \cite{Guck:Holm}.

The simplest example of this behavior is the rigid pendulum. If the system is given proper initial
conditions it approaches the position at which the pendulum points `upwards', spends some time near
it, and falls back to make a quick revolution across the lower part of its trajectory. The system
spends most of its time in the neighborhood of its top (unstable) equilibrium point, and can be
described statistically as being in equilibrium at that position, together with some correction for
the fast transit across the lower part of its orbit \cite{jimenez87}. The central role of unstable periodic
cycles has been emphasized in \cite{cvit:88}, and the correspondence with turbulent flows
has been explored in \cite{jim:kaw:sim:nag:shi:05,EckSchn:07}.

Well-known examples of turbulent structures are streaks and bursts in wall-bounded flows
\cite{kim:kli:rey:71,wal:eck:bro:72,lu:wil:73}, large-scale eddies in free-shear flows
\cite{brownr} and plumes in natural convection \cite{Pand:Schu:18}. All of them have been the
subject of thorough reviews that will not be repeated here, but the next two subsections will
briefly describe some salient features of the first two shear-driven cases, in part to point to open
problems, but also to illustrate more general questions about coherence that will be discussed
later. In these discussions, the three coordinate direction are $x,\, y$ and $z$, along the
streamwise, cross-shear and spanwise directions, respectively, and the corresponding velocity
components are $u,\, v$ and $w$. Capital letters are temporal averages, as in the mean velocity
profile, $U$, and lower-case letters are fluctuations with respect to these means. Primes
are reserved for root-mean-squared intensities.

\subsection{The coherent eddies of free-shear flows}\la{sec:FSF}
%
\begin{figure}
%
\centerline{%
\includegraphics[width=.65\textwidth]{\figpath 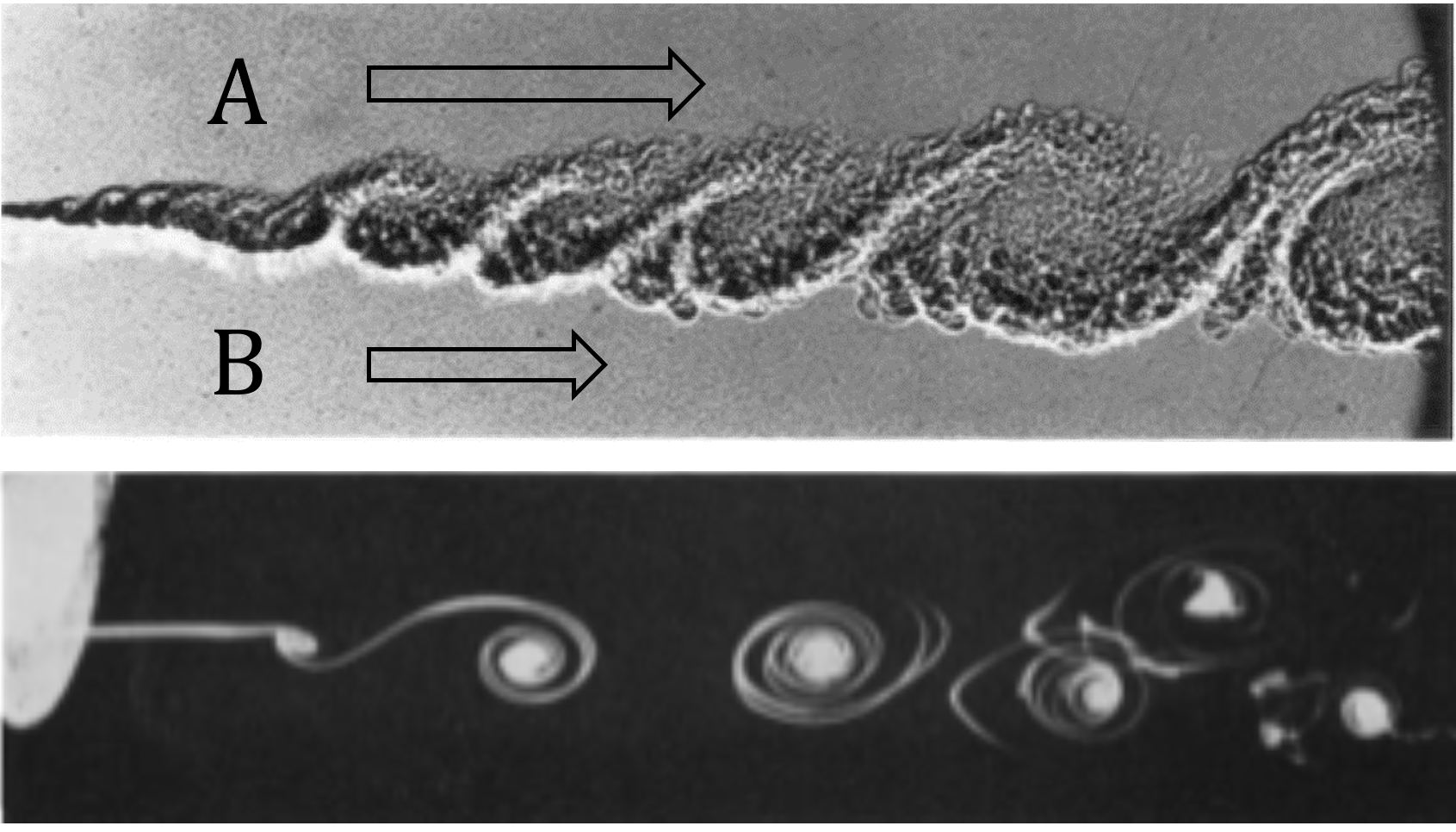}%
\mylab{.01\textwidth}{.265\textwidth}{(a)}%
\mylab{.01\textwidth}{.082\textwidth}{(b)}%
}
\caption{%
(a) Plane turbulent shear layer at high Reynolds number between two streams of different gases. Reproduced
with permission from \cite{brownr}. The Reynolds number based on the velocity difference and on the
maximum visual thickness is $Re\approx 2\times 10^5$.
(b) Initial development of a low-Reynolds-number velocity discontinuity. $Re\approx 7,500$.
Reproduced with permission from \cite{frey66}.
}
\la{fig:fsh3}
\end{figure}

Although evidence of organized motions in turbulence had been accumulating for some time, the first
clear demonstration of their existence were the shadowgraphs of a shear layer between
parallel streams in \cite{brownr} (Fig. \ref{fig:fsh3}a). They show that the interface between the two
streams forms large, organized, quasi-two-dimensional eddies that span the width of the layer and
persist when the Reynolds number is increased. One of the reasons why these pictures had a strong
influence in the field was that the origin of the coherent eddies was understood from the beginning. Very
similar structures at transitional Reynolds numbers had been published a few years before (Fig.
\ref{fig:fsh3}b), and the theory for the Kelvin--Helmholtz (KH) instability involved had been available
for over a century \cite{lamb,dra:rei:81}.

A shear layer forms when a low-speed stream, $U_B$, comes together with a higher-speed one,
$U_A=U_B+\Delta U$, at the end of a separating plate. It follows from dimensional considerations
that, if the Reynolds number is high enough for the effect of viscosity on the mean flow to be
neglected, the thickness of the layer grows linearly with the distance from the origin,
$x$ (Fig. \ref{fig:fsh2}a). This behavior is reproduced by relatively simple eddy-viscosity models
that essentially substitute turbulence by a homogeneous fluid with a modified viscosity, and, given
this success, the large-scale waves in Fig. \ref{fig:fsh3}(a) were a surprise. Especially, since
the obvious explanation was a linear theory developed for laminar flows.

Linear hydrodynamic stability is similar to that of other mechanical systems. Stable perturbations remain
small and die in the presence of friction, and unstable ones diverge from the equilibrium state.
But the concept needs some qualification when applied to turbulence, because the basic flow is not
steady, and the instabilities have to compete with other flow processes to be able to grow. Only
those with sufficiently fast growth rates are relevant.

\begin{figure}
\centerline{%
\raisebox{0mm}{\includegraphics[width=.40\textwidth]{\figpath 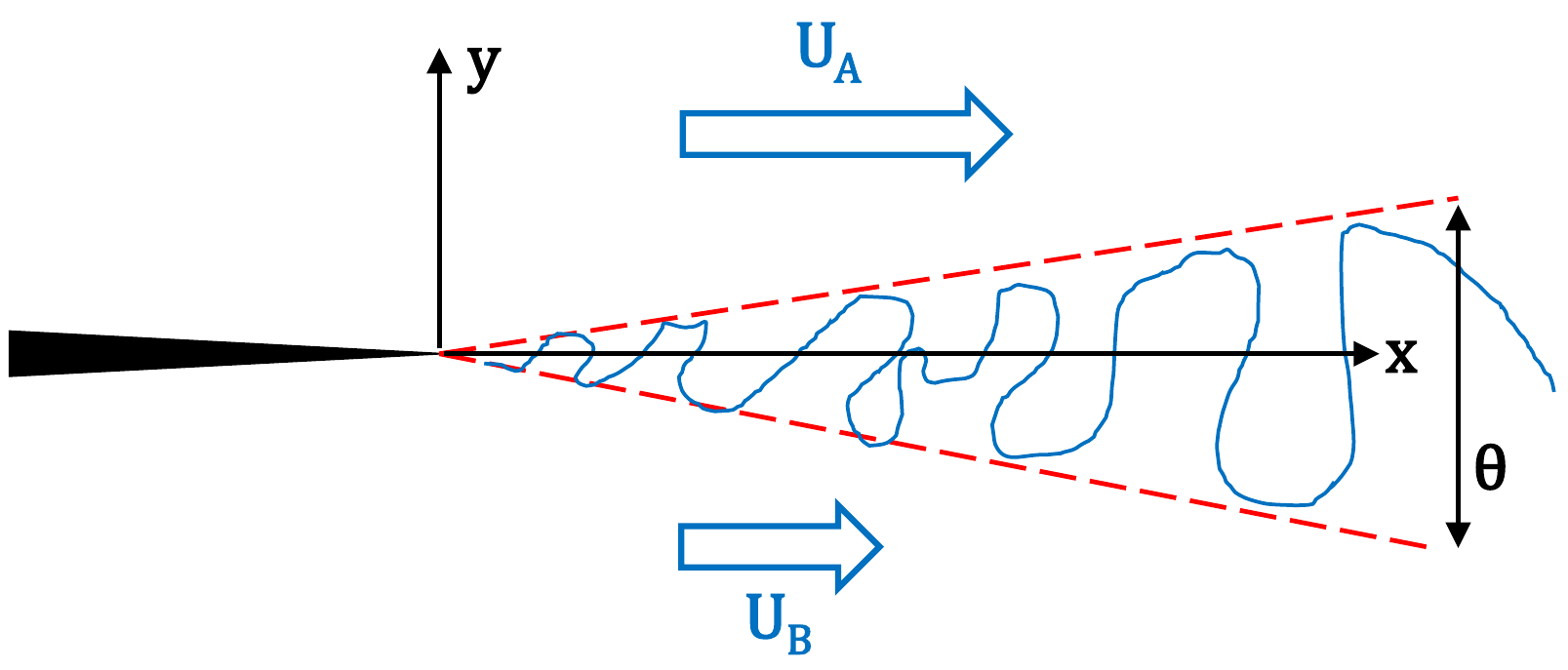}
\mylab{-.37\textwidth}{.12\textwidth}{(a)}}%
\hspace*{5mm}%
\raisebox{-4mm}{\includegraphics[width=.40\textwidth]{\figpath 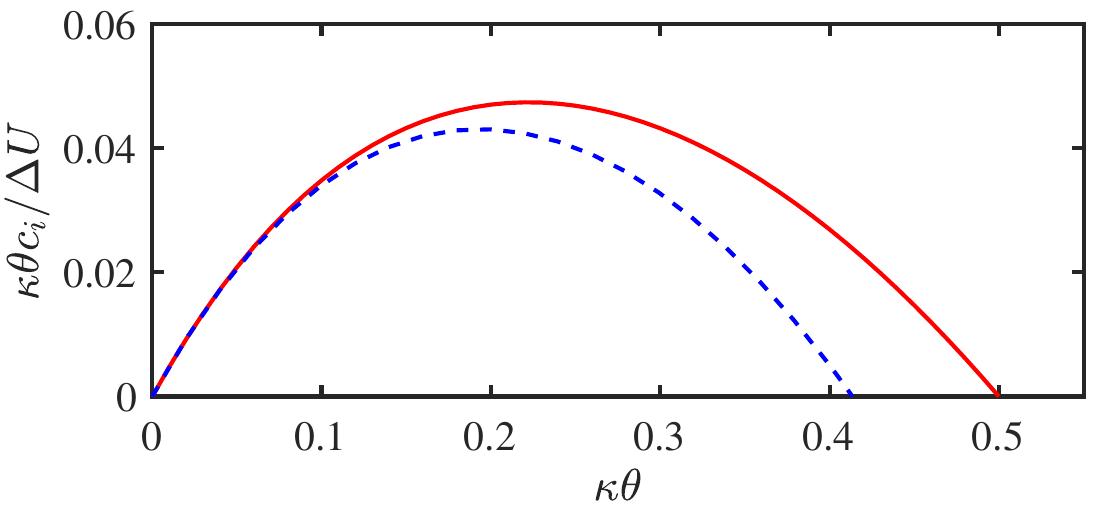}
\mylab{-.33\textwidth}{.14\textwidth}{(b)}}%
}
\caption{%
(a) Sketch of a turbulent shear layer.
(b) Growth rate, $\kappa c_i$, of the unstable modes for two model velocity profiles of the shear layer.
\solid, $U=\tanh(y)$; \dashed, $U={\rm erf}(y)$.
}
\la{fig:fsh2}
\end{figure}

Empirically, the spreading rate of shear layers is small, $\dr\theta/\dr x<0.05$, where
$\theta=\Delta U^{-2} \int (U_A-U)(U-U_B)\dd y$ is the momentum thickness of the layer, and the base
flow for the stability analysis can be approximated by a uniform infinite layer of vorticity between
two irrotational streams. The solution for the linearized perturbation equations can then be
expanded in Fourier harmonics, $q(x,y,t)=\sum \hu_\kappa(y) \exp \ii \kappa [x-c(\kappa) t]$, where $q$
stands for any of the velocity components. The perturbations are modally unstable whenever the
imaginary part of the growth rate is positive for some range of wavenumbers, $\kappa c_i>0$. The
results for two approximate mean velocity profiles, $U(y)$, are shown in Fig. \ref{fig:fsh2}(b), and
are functions of the dimensionless wavenumber, $\hkappa= \kappa \theta$. The KH instability is a
robust property of isolated vorticity layers, which works in spite of their possible imperfections,
and is essentially independent of the Reynolds number. As a consequence, vortex layers
are almost always linearly unstable, including turbulent ones such as the one in Fig.
\ref{fig:fsh3}(a). The resulting structures are typically large, because the KH instability is a
long-wavenumber phenomenon in which the vortex layer meanders as a unit. The minimum unstable
wavelength, $\lambda_s=2\pi/\kappa_s$, is given by the neutral limit in Fig. \ref{fig:fsh2}(b),
\beq
\hkappa_s \approx 0.45,
\quad\Rightarrow\quad
\lambda_s/\theta \approx 14.
\la{eq:oster}
\eeq
These large structures feel the effect of the smaller and faster ones as an eddy viscosity, which is different
from the molecular one, but which only has a minor effect on their behavior. The most characteristic
property of free-shear flows is that the linearized dynamics of their large scales is unstable.

That linear processes might be important in turbulence is not surprising. Turbulence is generally a
weak phenomenon. The root-mean-squared velocity fluctuations in Fig. \ref{fig:fsh3}(a) are small
compared with the velocity difference between the two streams $(u'/\Delta U\approx 0.15)$. The
internal deformation time of large eddies of size $\theta$ is $O(\theta/u')$, longer than the
shearing time of the mean flow, $\theta/\Delta U$. The fastest time controls dynamics, and the large
scales are dominated by the essentially linear processes driven by the mean flow. In fact, this is
typically how turbulence is maintained. The energy source of shear turbulence is the shear, and
energy enters the fluctuations when they interact with it. This single-scale interaction is linear,
but the analysis does not apply to the smaller flow scales. The deformation time of the turbulent
eddies, $\lambda/u_\lambda$, decreases with their size, and the linear processes that are important
for the larger scales do not necessarily remain relevant for the smaller ones.

\begin{figure}
\centerline{%
\includegraphics[width=.43\textwidth]{\figpath 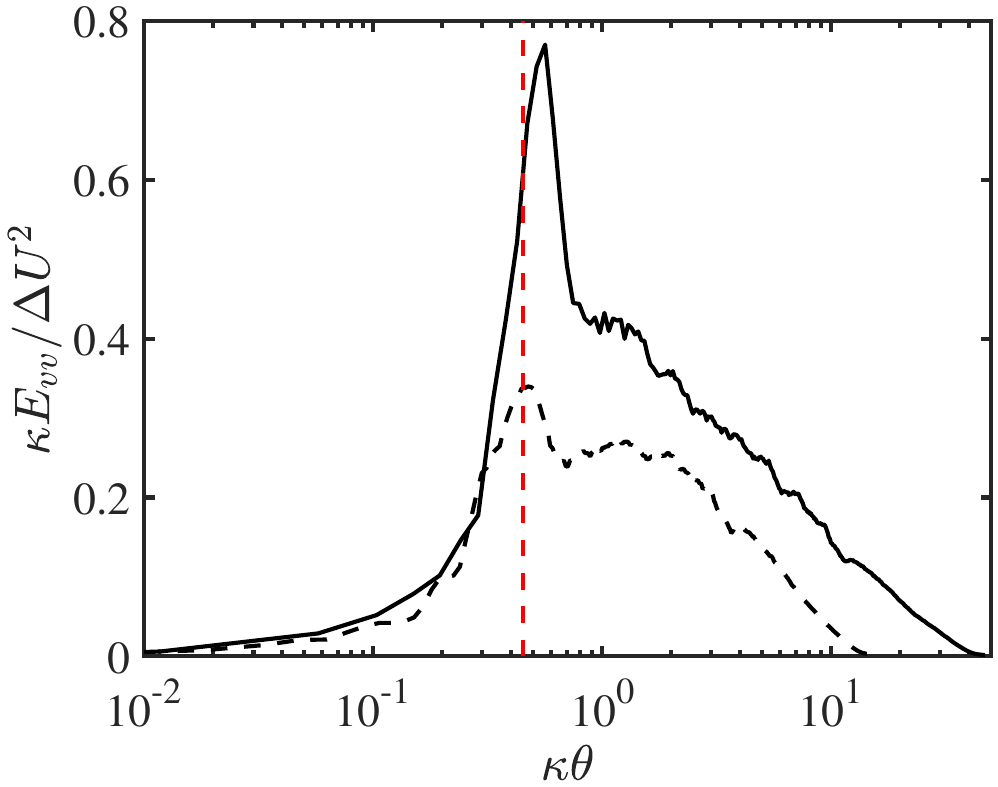}
\mylab{-.34\textwidth}{.28\textwidth}{(a)}%
\hspace{5mm}%
\raisebox{1mm}{\includegraphics[width=.435\textwidth]{\figpath 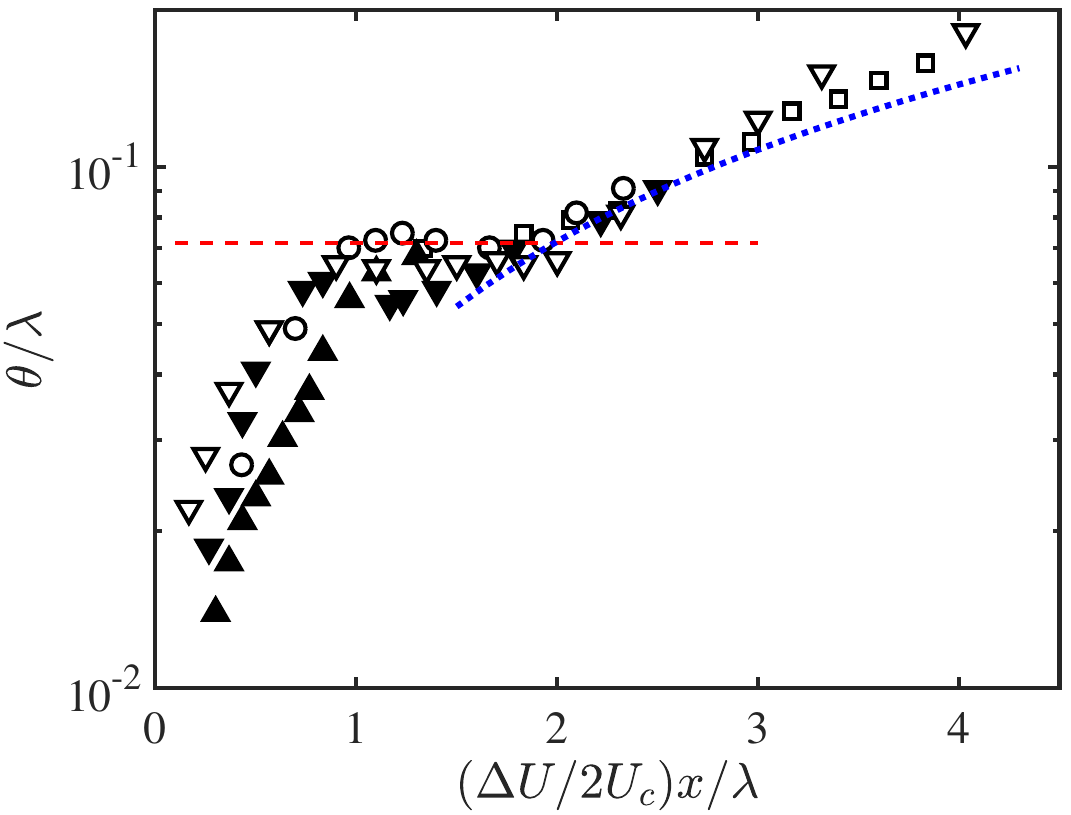}
\mylab{-.34\textwidth}{.28\textwidth}{(b)}}%
}
\caption{%
(a) Energy spectra of the transverse velocity component at the centerline of a free shear layer,
plotted against the wavenumber normalized with the local momentum thickness \cite{delville}. The
dashed vertical line is the limit of stability in Eq. \r{eq:oster}, $\kappa\theta=0.45$. The two
spectra correspond to different positions along the layer. \dashed, $\Delta U\theta/\nu = 3.9\times
10^3$; \solid, $1.2\times 10^4$.
(b) Growth of the momentum thickness of shear layers forced at the exit of the splitter plate
with perturbations of frequency $\lambda/U_c$. Data from various experimenters and velocity ratios
\cite{brow87}. The dashed horizontal line is Eq. \r{eq:oster}, and the dotted diagonal is the growth rate
of an unforced layer.
}
\la{fig:force_shl}
\end{figure}

It turns out that not only the existence of organized eddies in free-shear flows can be traced to
linear processes, but that many of their properties follow the predictions of the stability
analysis. Figure \ref{fig:force_shl}(a) shows spectra of the transverse velocity. The higher of the
two Reynolds numbers in the figure is comparable to the right-hand end of the shear layer in Fig.
\ref{fig:fsh3}(a), while the lower one is closer to its 10\% station. The dashed vertical line is
the stability limit in Eq. \r{eq:oster}, and it is clear that it divides the spectrum into two
distinct regions. In the unstable range to the left of the threshold, there is a roughly
exponential growth with wavenumber, while to the right there is a milder decay.

This behavior was explained in \cite{ho:hue:84}, who noted that, since the momentum thickness
$\theta$ is proportional to $x$, a wave with a fixed wavenumber $\kappa$ moves to the right in the
$\kappa\theta$ stability diagram as it travels downstream (Fig. \ref{fig:fsh2}b). As long as
$\kappa\theta$ is below the stability limit, the wave grows exponentially but, as soon as it is
carried beyond the threshold by the growth of the momentum thickness, it stops growing and the
regular turbulent dissipation damps it algebraically. Note that this means that the dominant waves
at each station are not those whose wavelengths are amplified faster, but those which have been
growing for a longer time. To the left of the stability threshold the structures are essentially
linear, but to its right the linear processes are not fast enough, and nonlinearity controls the
flow. This is where turbulence resides, and the separation underscores the second simplifying
feature of free-shear flows, which is that the energy-injection range of scales, to the left of the
stability threshold, is separate from turbulence itself, to its right.

Although not obvious from the previous discussion, it is interesting that, in this
approximation, each wave is only linearly amplified by a finite amount, and that the total
amplification factor is independent of the wavenumber. Consider in fact a profile for which the
imaginary part of the phase velocity is $c_i = \Delta U f(\hkappa)$, and assume that $\theta/x = C_L
\Delta U/U_c$, where $U_c=(U_A+U_B)/2$, and $C_L \approx 0.02$ experimentally. Since the real part
of the phase velocity of the unstable waves is constant, and approximately equal to $U_c$, their
wavenumber does not change as the waves move downstream, and their amplitude evolves as
\beq
\frac{1}{|\hu|}\, \frac{\dr |\hu|}{\dr t} =
\frac{U_c}{|\hu|}\, \frac{\dr |\hu|}{\dr x} =
\kappa \Delta U f(\hkappa),
\la{eq:ampli_2}
\eeq
This equation \r{eq:ampli_2}
can be recast in terms of $\hkappa$, which varies from $\hkappa=0$ at $x=0$ to
$\hkappa=\hkappa_s$ at the point where waves stop growing. Using the experimental values above,
\beq
C_L \frac{1}{|\hu|}\, \frac{\dr |\hu|}{\dr \hkappa} = f(\hkappa),
\quad\Rightarrow\quad
\frac{|\hu|(\hkappa_s)}{|\hu|(0)} = \exp\left(C_L^{-1}\int_0^{\hkappa_s} f(\xi)\dd\xi \right)
\approx 9,
\la{eq:ampli_3}
\eeq
which is reached after an evolution time $\Delta U\, t/\theta_s \approx 20$, where we have used Eq.
\r{eq:oster} to define the momentum thickness at the last unstable point. 

Equation \r{eq:ampli_3} also suggests that, when expressed in terms of $\hkappa$, the
velocity spectrum to the left of the stability threshold should be similar at all the streamwise
stations of the layer. This is approximately satisfied in Fig. \ref{fig:force_shl}(a), but the
details, and in particular the absolute amplitude, depend on the intensity $|\hu|(0)$ with which
each wavenumber is forced at the splitter plate. Note that the saturation of the growth of the perturbation
is a nonlinear effect, because the thickening of the layer is due to the quadratic Reynolds
stresses induced by the perturbations themselves. We will meet in the next section cases in which
the growth saturation is purely linear.

The quasi-linear character of the coherent structures soon suggested the possibility of manipulating
the layer by exciting or damping the initial perturbations, first for the purpose of research, and
later for control. As an example of a control application, the growth of the thickness of shear
layers in several experiments forced at the splitter plate is compiled in figure
\ref{fig:force_shl}(b). The initial growth of the shear layer depends of the forcing frequencies and
amplitudes but, as soon as the thickness becomes large enough for those initial waves to stop
amplifying, the growth ceases until other `natural' initial perturbations become strong enough to
resume normal growth. We can relate the thickness plateau to the wavelength $\lambda$ of the initial
forcing using the linear stability limit in Eq. \r{eq:oster}, which is the dashed horizontal line in
figure \ref{fig:force_shl}(b). Note that the perturbations in the figure only grow by an amount of
the order of Eq. \r{eq:ampli_3} from their amplitude at the exit of the splitter plate to their
saturation at $\hkappa=\hkappa_s$.

\subsection{Wall-bounded flows}\la{sec:orr}
   
The structures of free-shear flows had been basically understood by the end of the past century, and
are today part of the toolbox for manipulating these flows \cite{blooming03}. A modern review is
\cite{BrownRosh2012}. Our reason for including them here has been to show that some flows can be
described fairly completely in terms of coherent structures, and that this description is useful
both theoretically and in applications.

We next turn to wall-bounded flows, such as boundary layers, pipes and channels, in which coherence
has been harder to unravel, even if the evidence in them predates that in free-shear flows. We will
occasionally use `wall' units based on the kinematic viscosity, $\nu$, and on the friction velocity
$u_\tau$. The scale of the velocity fluctuations is everywhere $u_\tau$. The viscous length scale,
$\nu/u_\tau$, is important near the wall, while the flow thickness, $h$, characterizes the outer
flow. The range of length scales is therefore of the order of the friction Reynolds number, $Re_\tau
=u_\tau h/\nu$, which is typically $O(10^2-10^4)$ in simulations or experiments, and may be much
higher in industrial applications or in geophysical flows \cite{tenn,pope:00}.

The first indication of structure in these flows was the accumulation of tracers in boundary layers
into long streamwise-velocity streaks, subject to irregular cycles of break-up and regeneration
\cite{kli:rey:sch:run:67}. Those cycles were soon associated with intermittent bursts of the
wall-normal velocity \cite{kim:kli:rey:71}, classified into sweeps moving towards the wall, and
ejections moving away from it. While it was understood from the start that the bursts create the
streaks by `lifting-up' the mean velocity profile, the origin of the bursts was less clear and is
still a matter of some debate. A modern survey is \cite{jim18}.

It had originally been proposed that wall-bounded turbulence is maintained by the marginal modal
instability of the mean velocity profile \cite{malkus56}, but this was found to be incorrect in
\cite{reytied67}, and the argument moved to whether an instability of the streaks could be
responsible for the regeneration of the bursts. Such instabilities are not hard to find, because a
streamwise-velocity streak is bounded by sidewalls of intense wall-normal vorticity that may become
subject to KH instability \cite{swe:black:87}. Unfortunately, the details proved difficult to pin
down in realistic models, because the wall-normal shear of the mean profile tends to inhibit the KH
structures of the sidewalls.

A breakthrough was the realization in the 1990s that the growth of perturbations does not always
require exponentially growing modal instabilities \cite{tref_etal_93,schmid01,schm07}, and that,
even if these non-modal instabilities are usually transient, they may evolve into more permanent
nonlinear structures. It was found, for example, that the most transiently amplified perturbations
of the mean velocity profile of turbulent channels, once corrected by limiting their growth time or
by an eddy viscosity, are very similar to the streaks and bursts found in real flows
\cite{but:far:93,ala:jim:06}.

There are at least two transient amplification mechanisms that are important for wall-bounded
turbulence. The simplest is the lift-up mentioned above, in which a persistent perturbation of
the wall-normal velocity deforms the mean shear to generate a perturbation of the streamwise
velocity, which grows linearly in time,
\beq
\p_t u = -v \p_y U + \ldots
\quad\Rightarrow\quad
u\approx -\p_yU\int v\dd t.
\la{eq:liftup}
\eeq
What makes this mechanism transient is the lack of feedback on $v$, which has to be independently
maintained to keep $u$ growing. The second transient mechanism was proposed by \citet{orr07a} in the
early twentieth century, and has been rediscovered and extended several times since then
\cite{far_ioa93a,jim:13a}. Essentially, when a localized perturbation of $v$ interacts with a shear, it is
progressively tilted forwards by it. If the initial configuration is tilted backwards, the shear
makes it more vertical, and the perturbation is algebraically amplified because $v$ has more space
to act, but once it passes the vertical position and begins to be tilted forwards, it is damped
again, because the wall-normal velocity is forced to work against itself. If the perturbation is
uniform along the span, its velocity is eventually fully damped, but oblique or span-localized perturbations
leave a residue in the form of a wake in $u$ \cite{jim:13a}. The lift-up and the Orr mechanism draw
energy from the mean shear into the turbulent fluctuations. A variant of the Orr process acts on the
spanwise shear in the sidewalls of the streaks, $\p_z u$, and has recently become known as
`push-over' \cite{loz:etal:21}, although it is probably equivalent to the non-modal streak
instability proposed in \cite{Schoppa02}. This mechanism isotropizes the energy of the streak.

That the growth of the perturbations is transient should not be considered a problem. Recall that
the nominally exponential modal KH instability of free-shear layers is also limited to a modest
growth over a few shear times. The saturation mechanism is in that case nonlinear, while that of the
Orr mechanism is linear, but the time scale is controlled in both cases by the shear. The precise
limit depends on how the shear is defined, especially in the very inhomogeneous profiles of
wall-bounded turbulence, but it is usually $(\p_y U)\, t=O(10)$ \cite{jim:13a}. The maximum
amplification is also $O(10)$. The most important effect of the Orr mechanism in shear flows is not
to generate growth, which depends on the initial conditions, but to define the inverse of the shear
as the time in which fluctuations are damped, and thus lose memory of their past evolution.

All the processes just described are known to be important for the maintenance of wall-bounded
turbulence, in the sense that if the associated terms of the Navier--Stokes equations are modified,
turbulence is strongly affected and often decays \cite{jim:pin:99,loz:etal:21}. Even before those
conceptual experiments had been carried out, visual inspection of numerical simulations and
visualization experiments had converged to a regeneration cycle in which bursts create the streaks,
and some kind of streak instability generates the bursts \cite{jim:moi:91,Hamilton95,jim:pin:99}.
Many of the articles mentioned above include characterizations of this cycle.

However, there are problems with this simple picture. In the first place, while free-shear
turbulence is a single-scale flow from the point of view of energy production, wall-bounded
turbulence is not. The length scale of the Orr process increases proportionally to the distance from
the wall, spanning a factor that we saw at the beginning of this section to be $O(Re_\tau)$.
In this range of scales, the production of energy  coexists with its cascade towards dissipation.
Moreover, since multiscale flow fields are difficult to interpret, efforts are often made to
simplify the problem as much as possible before analyzing it, and most characterizations of the
turbulence cycle have been based on minimal flow units \cite{jim:moi:91,flo:jim:10}, which are
designed to contain a single set of structures, and which therefore miss multiscale interactions.

\begin{figure}
\centerline{%
\raisebox{0mm}{\includegraphics[width=.42\textwidth]{\figpath 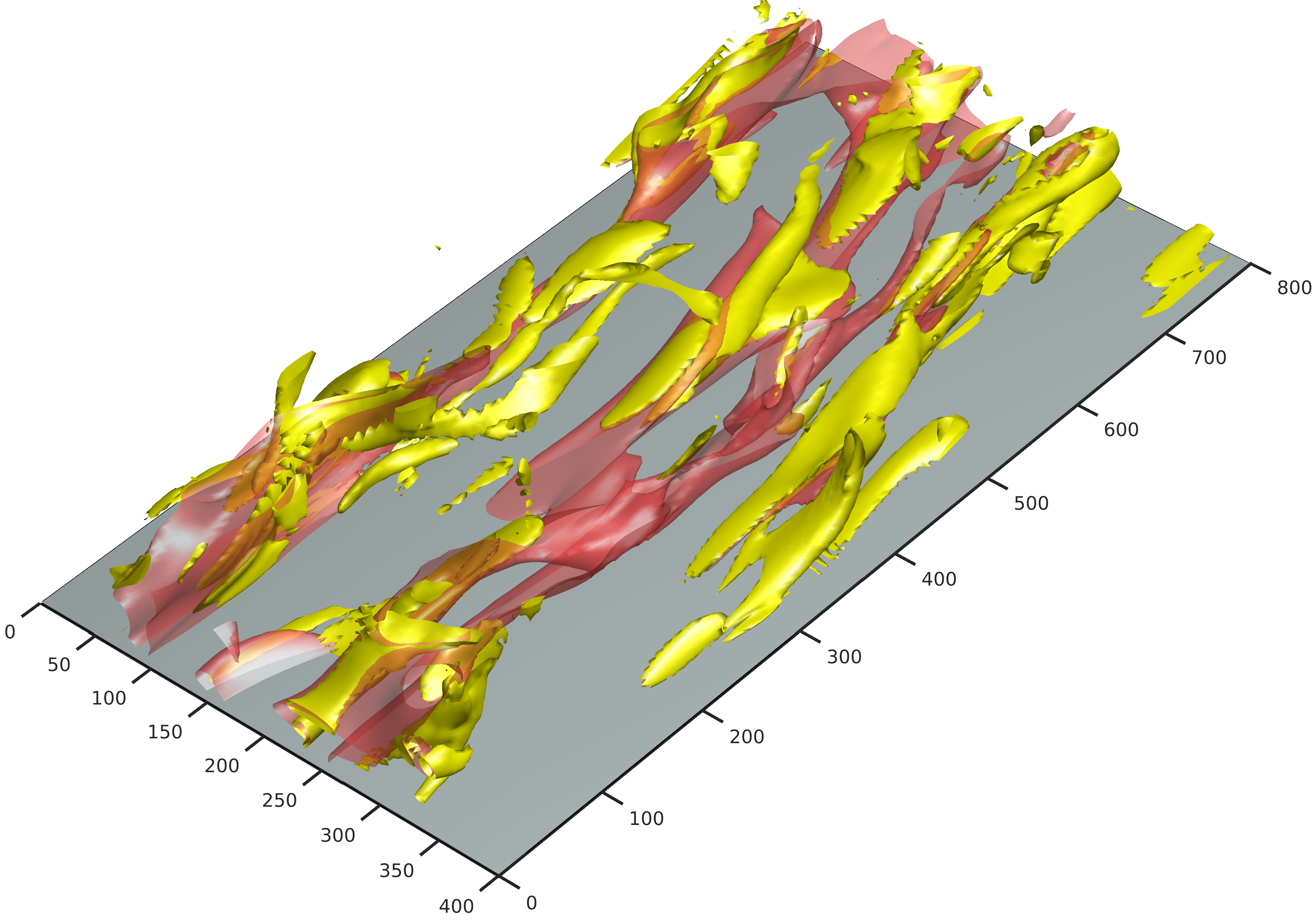}%
\mylab{-.35\textwidth}{.22\textwidth}{(a)}}%
\hspace*{2mm}%
\raisebox{0mm}{\includegraphics[width=.43\textwidth]{\figpath 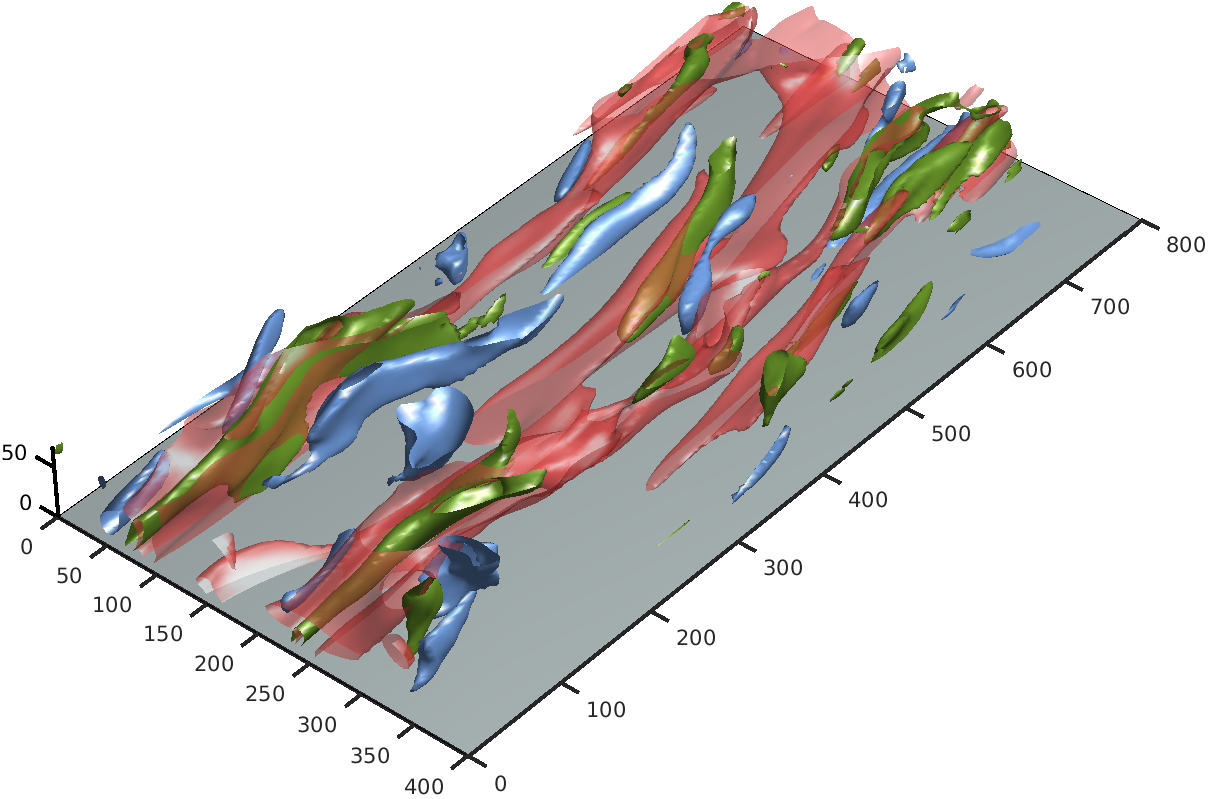}%
\mylab{-.35\textwidth}{.22\textwidth}{(b)}}%
}%
\vspace{0mm}%
\centerline{%
\raisebox{0mm}{\includegraphics[width=.42\textwidth]{\figpath 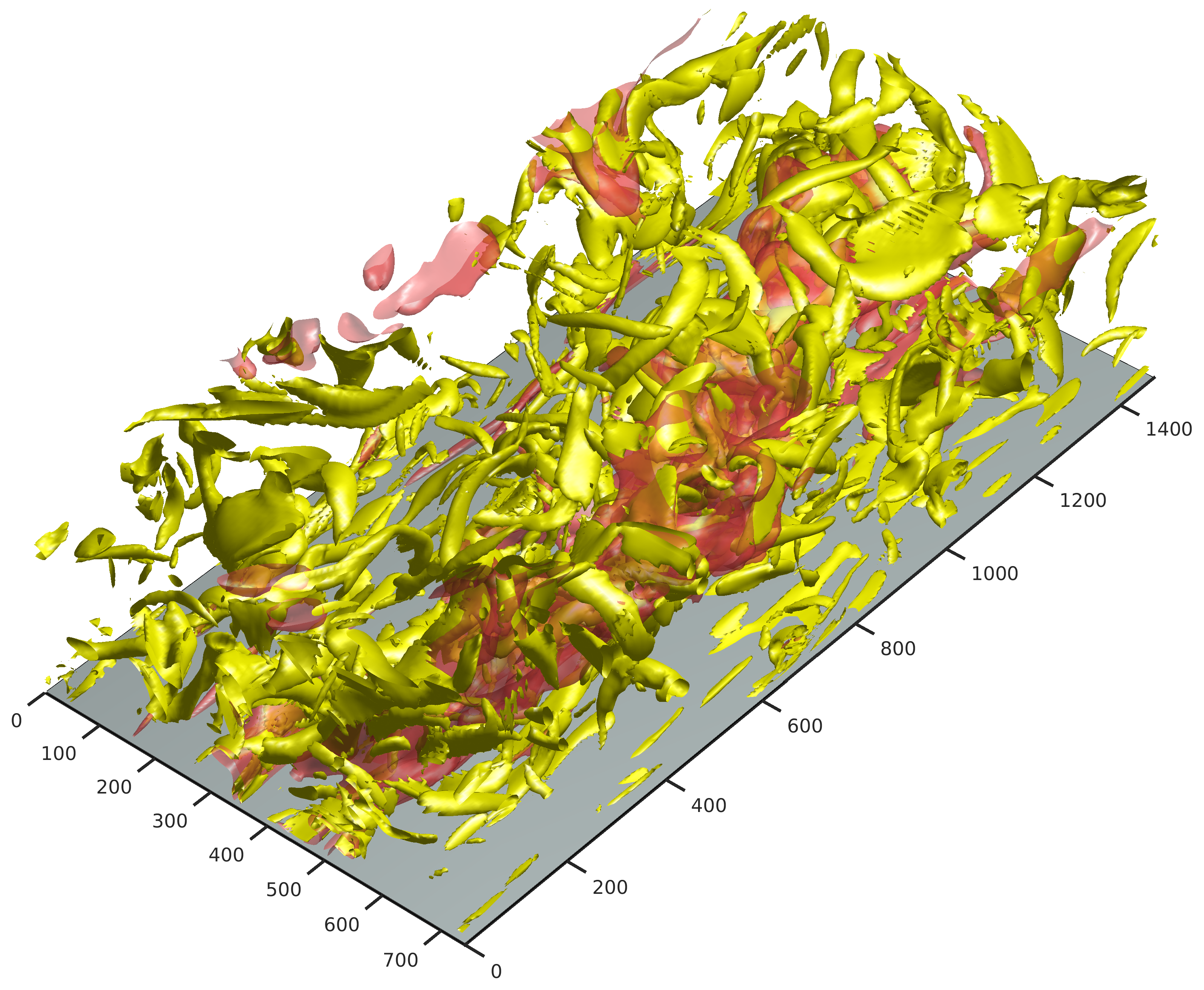}%
\mylab{-.35\textwidth}{.26\textwidth}{(c)}}%
\hspace*{2mm}%
\raisebox{0mm}{\includegraphics[width=.43\textwidth]{\figpath 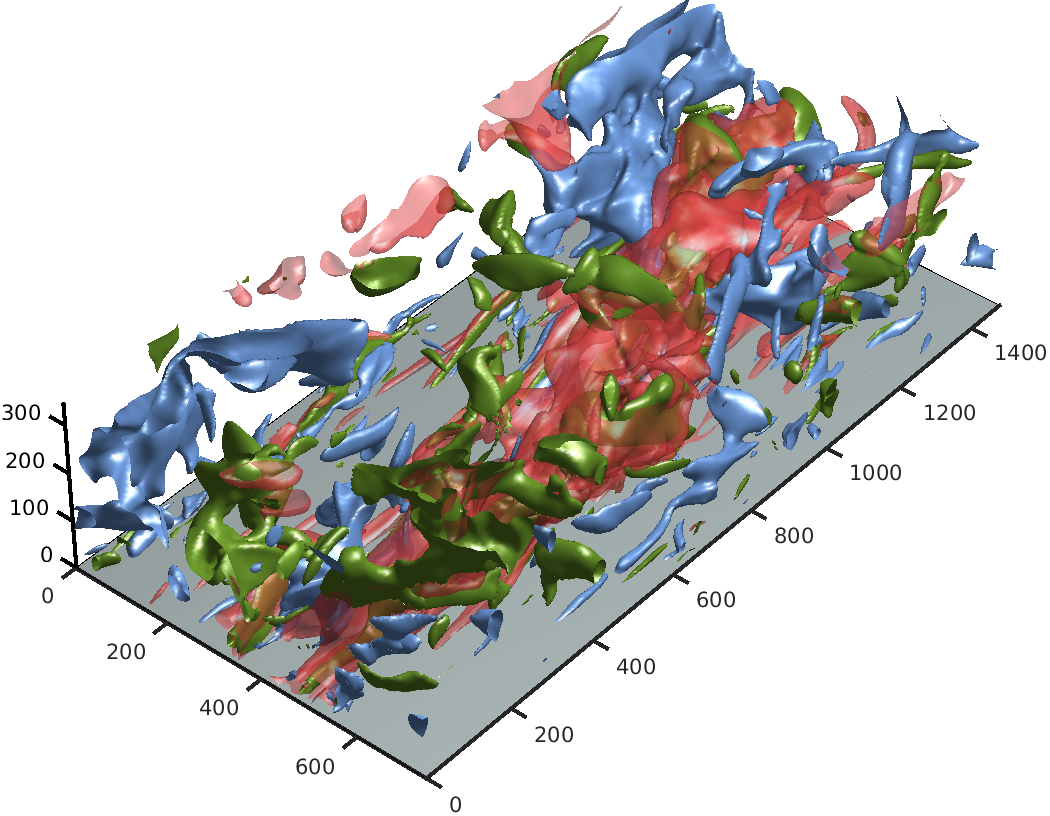}%
\mylab{-.35\textwidth}{.26\textwidth}{(d)}}%
}%
%
\caption{%
Structures of a turbulent channel in a small computational box, $(L_x\times L_z)=(\pi/2\times
\pi/4)h$, $Re_\tau=940$. The four panels are the same flow field, flowing from bottom left to top right. 
(a,b) For $y\utau/\nu\le 70$. (c,d) $y\utau/\nu\le
350$.
The translucent red objects in all panels are low-velocity streaks, $u<-1.3\, u'(y)$. 
The yellow objects in (a,c) are vortices defined as $|\omega|<1.75\, |\omega|'(y)$.  
The green and blue objects in (b,d) are, respectively, positive and negative $v$-structures,
defined as $|v|<1.75\, v'(y)$.
Panels (c,d) include the full wall-parallel domain, while (a,b) only include one quarter of the
plane, approximately centered spanwise on the streak in (c,d) at $x=0$. Axes are in wall units.
}
\la{fig:streaks}
\end{figure}

\begin{figure}
\centerline{%
\raisebox{0mm}{\includegraphics[width=.80\textwidth]{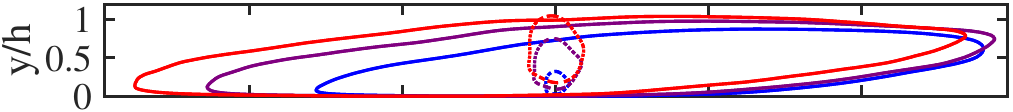}
\mylab{-.69\textwidth}{.06\textwidth}{(a)}}%
}%
\vspace*{1mm}%
\centerline{%
\raisebox{0mm}{\includegraphics[width=.795\textwidth]{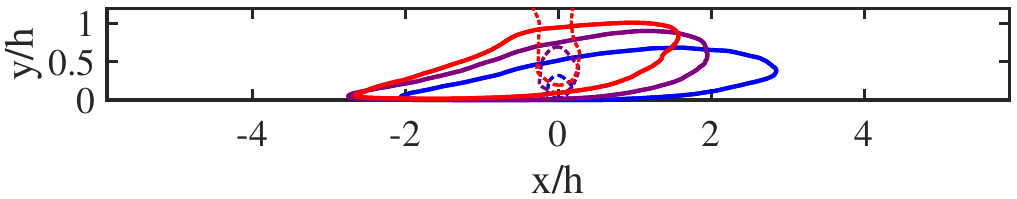}
\mylab{-.69\textwidth}{.12\textwidth}{(b)}}%
}%
\caption{%
Two-point correlations of: \solid, streamwise velocity; \dashed, wall-normal velocity. Contour is $C_{**}=0.1$,
and correlations are centered at $y/h=0.1,\,0.3$ and 0.5.
(a) Channel at $\retau=2003$ \cite{hoy:jim:06}. 
(b) Boundary layer at $\retau\approx 1530$ \cite{sil:jim:mos:14}.
}
\la{fig:corrcuts}
\end{figure}

One consequence is that our view of the structures of the turbulence cycle has evolved with
the Reynolds number of the available simulations. Early descriptions were motivated by 
simulations at low Reynolds numbers, and only included velocity streaks and quasi-streamwise vortices,
with the latter being responsible for the fluctuations of $v$. These early simulations contained
little beyond the viscosity-dominated buffer region, where velocity and vorticity share the same
length scale, but this is not true farther from the wall.
 
The difference can be seen in Fig. \ref{fig:streaks}. The four panels correspond to the same flow
field, but the top two panels are truncated above the buffer layer, while the bottom two extend to
the lower part of the logarithmic layer. The domain of Figs. \ref{fig:streaks}(a,b) is
approximately one quarter of that of Figs. \ref{fig:streaks}(c,d), for clarity, but the structures
in the former are otherwise the roots of those in the latter. The red low-velocity streaks and the
quasi-streamwise yellow vortices in Fig. \ref{fig:streaks}(a) have similar sizes, and it is natural
to assume that they directly interact with one another. Fig. \ref{fig:streaks}(b) also shows the red
streaks, now accompanied by green and blue structures of the wall-normal velocity, which also have
scales similar to those of the streaks and vortices in Fig. \ref{fig:streaks}(a). However, the
scales of the two kinds of structures are very different in Figs. \ref{fig:streaks}(c, d). The
vortices in Fig. \ref{fig:streaks}(c) are only slightly thicker than those in Fig.
\ref{fig:streaks}(a), but the streak is much wider, $O(h)$, and the regions of strong wall-normal
velocities in Fig. \ref{fig:streaks}(d) are considerably shorter than those of $u$. The orientation
of the different structures is also now very different.

Reduced models of the minimal unit of the buffer layer have been quite successful \cite{Waleffe97},
and exact permanent-wave and periodic-orbit solutions of the Navier--Stokes equations are available.
They look extraordinarily like Figs. \ref{fig:streaks}(a, b)
\cite{Nagata90,Waleffe2001,gibson:08,KawEtal12}, including a streak and two counterrotating
vortices, and are the best available candidates for the coherent invariant sets discussed at the
beginning of \S\ref{sec:coherent} \cite{jim:kaw:sim:nag:shi:05}.

In contrast, it is difficult to think of a model in which the large velocity streak in Figs.
\ref{fig:streaks}(c, d) is sustained directly by the disordered small vortices in the figure, although
large-scale streaks are known to undergo an intermittent decay and regeneration cycle very similar to those
in the buffer layer \cite{flo:jim:10}, and large-eddy invariant solutions in a uniform shear were
presented in \cite{sek:jim:17}. These solutions are linked to the shear, not to the presence of a 
wall. Similar cycles are known for virtually all shear flows in which the mean velocity
profile is not subject to modal instabilities \cite{jim18}.

Unfortunately, minimal flows invite misinterpretations. For example, the correlations in Fig. \ref{fig:corrcuts}
show that the streamwise velocity is much longer than the wall-normal velocity
\cite{sil:jim:mos:14,jim18}, and minimal flows, which are designed to contain a single copy of the
$v$-structures, see streaks as being infinitely long. This has led some models to include 
infinitely long streaks as a required ingredient \cite{far_ioa12,thom:etal:15}, but direct
experimentation shows that  streaks only need to be about twice as long as the bursts to
sustain turbulence \cite{jim22_nostr}. 

This, of course, does not mean that very long streaks do not exist in natural flows, but it opens
the fascinating possibility that there may be several turbulence regeneration mechanisms, one of
which contains a strong streak while others do not. If that were the case, it is unclear whether one
of them dominates the others, or whether they coexist and, if they do, whether they alternate in
time or space, or somehow share the same flow regions. Flows in which long streaks have been
inhibited look different from those with a streak, with a striking diagonal structure
\cite{jim22_nostr} that is also weakly present in natural boundary layers and channels
\cite{sil:jim:mos:14}. Moreover, streaks are not universal. The correlations of $u$ in the channel
in Fig. \ref{fig:corrcuts}(a) are longer than those in the boundary layer in Fig.
\ref{fig:corrcuts}(b), even if the correlations of $v$ are similar in the two cases, and the
$u$-structures of Couette flow are believed to be essentially infinitely long \cite{lee:moser:18}.
That streaks in natural flows are so much longer than the bursts from which they are
supposedly generated seems to imply that the smaller structures of $v$ align with one another at
the larger scale of $u$, but the way this is implemented in the flow remains unclear.

Compared to free-shear flows, it is harder to find control strategies motivated by dynamics in
wall-bounded flows. In part, this is because we understand wall turbulence less well than free shear
flows, but also because the inhomogeneity of wall turbulence requires very fast and very small
control actuators, which are impractical. The obvious place to actuate on a wall-bounded flow is the
wall, but the time and length scales are there much smaller than in the boundary layer as a whole.
For example, the viscous regeneration cycle mentioned above suggests that actively damping the
quasi-streamwise vortices should damp turbulence. This is confirmed by simulations \cite{choi94},
but the length scale of the actuation is $\nu/\utau=h/Re_\tau$, which is typically microns in
realistic flows. Similarly, the time scale of the actuation is determined by the shear at the wall, $1/S_{w}
\sim Re_\tau^{-1} h/u_\tau $, which is of the order of hundreds of KHz for a typical airplane wing.
A more practical approach that works passively on similar principles are riblets
\cite{Walsh84,Luchini91}, but their length scale is still of the order of microns.

A further problem with controlling wall-bounded turbulence are instabilities. We saw in
\S\ref{sec:FSF} that the KH instability is a property of layers of intense vorticity, such as the
near-wall viscous layer. The instability of this wall layer is inhibited by impermeability,
but control manipulations can excite it and degrade the control efficiency
\cite{jim:uhl:pin:kaw:01,gmayjim11}. Similarly, attempts to bypass the small near-wall scales by
acting directly on the large-scale regeneration cycle of the core flow often run into instability problems 
\cite{guseva22}.


\section{The rest of turbulence}\la{sec:incoherent}
\subsection{Cascades}\la{sec:cascade}

The turbulence cascade does not properly belong to a paper on coherent structures in chaotic flows,
although the reason may be that we don't know enough to link the two, but we cannot completely avoid
the subject because we have found several examples of multiscale turbulence, and many of the open
questions that we have identified involve multiscale interactions. Cascades are models for the
those interactions. They do not necessarily imply the flow of some quantity
in scale space, but the most successful predictions are about those that do, because the
transferred quantity provides a link among the cascade stages.

The most familiar example of cascade was proposed by \citet{rich22} in a well-known stanza in which
eddies incrementally transfer their `velocity' down to the smallest viscous scales, by which he
probably meant their energy. This was quantified by \citet{kol41}, who computed the correct spectral
slope by recognizing the energy transfer rate as the cascading quantity. He did not explicitly
invoke a hierarchy of cascading eddies, but this was done by \citet{obu41}, who linked the
statistical argument in \cite{kol41} with the conceptual model in \cite{rich22}. Interestingly,
evidence for an eddy cascade in isotropic turbulence, local both in scale and in physical space, has
only recently been forthcoming \cite{cardesa17}, although relying on a eddy model based 
on kinetic energy, with little geometric or dynamical content. 
In fact, the primary reasons why we have not discussed the isotropic energy cascade up to now, even
if such a cascade is active below the energy-injection range in all three-dimensional turbulent
flows and is crucial to turbulent dissipation, is that we know relatively little about its coherent
structures.

The cascades in shear flows are better understood. As a first example, the limiting size that we
found for the instability of the eddies in free-shear flows suggests that something else has to take
their place after they stop growing. The answer in this case is that either neighboring eddies pair
into a new eddy twice as large \cite{brownr, win:bro:74}, or that an eddy is torn by its two
neighbors, which grow at its expense \cite{moo:saff:75}. Both mechanisms coexist in experiments
\cite{her:jim:82}, and the self-similar stability limit keeps the eddy size proportional to the
layer thickness.

We also met at least two multiscale processes in wall-bounded turbulence that
involve structures and require explanation: that the size of the bursts is proportional to their mean
distance to the wall, and that the streaks are much longer than the bursts.

The first of these observations was addressed in Townsend's attached-eddy model \cite{towns}, which
was cast in spectral terms in \cite{per:abe:75}, although, as in the case of \cite{kol41}, without a
specific eddy structure in mind. A kinematic model in terms of hairpin eddies was later developed in
\cite{per:hen:cho:86} and has become popular \cite{adr07}, although it is still unclear which are
the inter-scale interactions involved. The original spectral argument implies a momentum transfer in
which each cascade stage carries the full Reynolds stress at one distance from the wall but, as in
most cascade arguments up to now, how this information is shared among stages is not specified. The
momentum argument fails beyond lengths of $O(h)$, which is the longest expected size of the active
bursts, and the second question above, how streaks become as long as they are, remains unanswered.

\begin{figure}
\vspace*{.03\textwidth}%
\centerline{%
\raisebox{0mm}{\includegraphics[width=.92\textwidth,clip]{\figpath 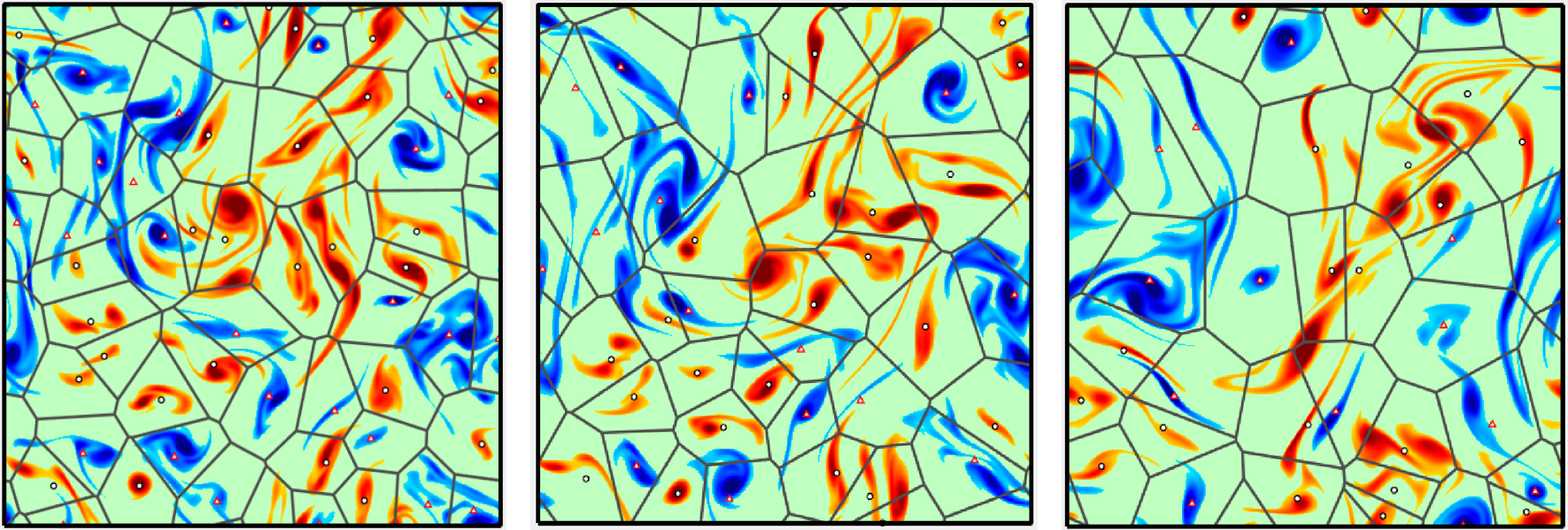}}%
\mylab{-.79\textwidth}{.33\textwidth}{(a)}%
\mylab{-.48\textwidth}{.33\textwidth}{(b)}%
\mylab{-.17\textwidth}{.33\textwidth}{(c)}%
}%

\caption{%
Three moments of the inverse energy cascade during the decay of a two-dimensional turbulence box.
Only vortices larger than some fraction of the mean vortex area are plotted, and the lines are the
Voronoi tessellation of their centers of mass. (a) $\omega'_0 t=1$. (b) $\omega'_0 t=3$. (c)
$\omega'_0 t=7$, where $\omega'_0$ is the rms vorticity at $t=0$. Note the coarsening of the cluster
of blue vortices on the left part of the flow.
\href{https://torroja.dmt.upm.es/~jimenez/2Dturbulencecascade.avi}{A video of this decay} can be
found in the Supplementary material.
}
\label{fig:cascade2d}
\end{figure}

All the cascades just mentioned are one-directional, in the sense that the size of the eddies
either decreases (direct) or increases (inverse) but, whenever interactions have been followed in
detail, both directions coexist. Either the resulting direction is the average of the two processes,
or the two cascades act on different quantities and proceed quasi-independently. For example, the
pairing process in free-shear flows is an inverse cascade towards larger eddies, but there is a
simultaneous cascade towards smaller scales, which  initially involves
quasi-streamwise vortices \cite{Bern:Rosh:86}.

The hierarchy of attached eddies in wall-bounded turbulence has also been mapped in some detail and
involves both eddy mergers and splits \cite{loz:jim:14}. The isotropic energy cascade also works
both ways, and the dominance of the direct energy flux is driven by probability arguments
\cite{ors77,cardesa17,VelaJim21}. In these two cases, the reasons why splitting dominates merging
are not hard to rationalize for individual eddies, but how the cascade adjusts itself to a power law
consistent with similarity has been harder to disentangle.

Particularly interesting is the case of two-dimensional turbulence, where simulations are relatively
inexpensive and easy to analyze. A tendency to form large structures was predicted by
\citet{onsag49} using equilibrium arguments on an energy-conserving Hamiltonian approximation, and
a dimensionally predicted inverse energy cascade has been well documented 
\cite{Boff:Eck:12}. The vorticity in two-dimensional turbulence quickly condenses into discrete
vortices that provide the elements for the Hamiltonian approximation in \cite{onsag49}. The inverse
cascade works by the successive coalescence of these vortices \cite{Benzi92} but, as above, how the
spectral exponent is implemented is unclear. Figure \ref{fig:cascade2d} suggests a possible
mechanism, and a short movie can be found in the supplementary material. The Voronoi polygons in the
figure define the `area of influence' of each vortex, which merge as the vortices do. Coalescence is
local to each vortex pair, while the spectrum describes a global arrangement, but the vortices
quickly rearrange themselves in a relatively uniform pattern that minimizes energy after each merger
\cite{jimenez21}. This appears to be how the spectrum is enforced and, together with the connection
of the original prediction with statistical mechanics, suggests that this particular inverse cascade is a
succession of equilibrium states, much closer to the coarsening of a foam than to a concatenation of
dynamic cascade steps. It is unknown whether a similar description applies to any of the other
inverse cascades mentioned above.

\subsection{The river}\la{sec:river}

The coherent structures discussed up to now are well-known examples from real flows, all of which
were discovered in experiments or simulations because they are intense enough to be distinguishable
from the rest of the flow. It makes sense that being strong helps coherence, but it is less clear
whether the quantities that are easy to observe are the only relevant ones for coherence, or whether
there are weaker, dynamically important, but unobserved structures. For example, the sweeps and
ejections that form the bursts discussed in \S\ref{sec:orr} explain 60\% of the Reynolds stresses of
wall-bounded flows, while only covering 8\% of the volume \cite{loz:flo:jim:12}. This is impressive, but
begs the question of what is happening within the remaining 92\% of the volume.

It may be useful at this point to reflect on a system in which we can identify both coherent
structures and an apparently incoherent background. Consider the water cycle on Earth. If we recall
our original definition of coherent structures as submanifolds whose dynamics is relatively
independent from the rest of the system, rivers satisfy it well: water that rains in the mountain
collects into valleys, runs downhill, and eventually reaches the ocean. The process is fairly
independent of other parts of the cycle, and even of the properties of water. There are
ethane--methane rivers in Titan \cite{pogg:etal:2016}, which are believed to behave approximately
like ours, even if, in the absence of a global ocean, other parts of their cycle are probably very
different from those on Earth.

The dynamics of rivers is relatively well understood: the evolution of a fluid particle is
transient, driven by gravity and with a lifetime controlled by topography. In addition, the system
of tributaries is a relatively good example of a self-similar cascade. Even so, rivers are not the
full water cycle, and they would stop running if the water that reaches the ocean did not return to
the head of the valleys. The return mechanism is not formed by rivers, and it is driven by the Sun,
rather than by gravity, but its details are unimportant to the river. The Amazon at the Equator
works essentially as the Yukon near the Arctic. The ocean, the clouds, and most of
geophysics, are unstructured from the point of view of rivers.

At least, this is the point of view of hydrologists and of dwellers in river valleys, but
oceanographers probably think that oceanic currents and the polar ice caps are also coherent parts
of the water cycle, and meteorologists may consider tropical cyclones and the jet stream as the
structures most relevant to water circulation. Both are probably right from the systemic point of
view. To them, rivers are an unstructured component of the cycle, and which river contributes which
water to a particular oceanic stream or to a given atmospheric storm is irrelevant. Identifying
coherence is relatively straightforward using the approximate-independence criterion outlined above,
but whether a structure is incoherent depends on which aspect of the phenomenon we are interested
in.

\subsection{Junk turbulence}\la{sec:junk}

The example of the river raises the question of whether the turbulence regeneration cycle that we
have described above is complete, or whether we should also look outside the catalog of known
shear-driven instabilities. There are numerous examples in other branches of science in which an
apparently useless part of the system turns out to have a crucial role. The best known example is
`junk DNA', which was considered for a long time to be unrelated to genetics because it does not
appear to code a protein, but which was eventually found to have a variety of functions, although
different from those of classical genes \cite{junk:14}.

It is natural to wonder whether something similar may be happening in turbulence, and whether we are
missing important flow elements because they are unrelated to the energy-momentum or vorticity
tensors that we usually observe. The situation is somewhat different from genetics, where everything
in the cell has been subject to natural selection for untold years, which tends to delete
useless features. The recent success of automatic prediction of protein folding patterns is apparently not
due to an exhaustive search over all possibilities, but to the exploitation of the relatively few
patterns present in nature \cite{AlphaFold3}.

However, there is no obvious selection mechanism to delete useless parts of a turbulent flow, and it
is possible that most of what we have not identified as important by now is really unrelated to
the turbulence generation mechanism. In fact, there are indications that this may be the case. As a
first example, bursts were originally identified as intense regions of the tangential Reynolds
stress, which is a momentum flux. But the quantity in the momentum equation is not the flux tensor,
but its divergence, and optimum fluxes can be found with the same divergence but with substantially
lower variability \cite{jim16}. In this sense, classical sweeps and ejections are largely redundant because
they label fluxes that eventually cancel among themselves without affecting the flow but, when
the structures of intense optimum fluxes are computed, they tend to be smoothed versions of 
Reynolds-stress bursts collocated with them \cite{osawa18}.

The previous example shows that structures identified by one particular intense property persist even
after most of their `junk' component is eliminated. \citet{osawa24} took the more direct approach of
identifying important flow features by their effects, rather than by their properties. They
perturbed random regions of the flow and observed the resulting changes after some time. The result
is again that significant and irrelevant regions are collocated with sweeps and ejections,
respectively.
 
Neither of these examples proves that sweeps and ejections are the only, or even the main,
dynamically significant features of shear flows, and the issue remains open to further
research, but the fact that two attempts to find something different ended up with the same
structures as the classical approach, suggests that we may have found the most important
contributors to the turbulence cycle.


\section{New techniques and a possible future}\la{sec:future}

It should be clear from our discussion that the study of turbulence has changed considerably
over the past fifty years. As in most scientific endeavors, the reason is unlikely to be that we are
smarter than our supporting giants, but rather that we have had access to better data and to new
theoretical frameworks. With few exceptions, observational turbulence in the 1950's was treated  as
a random process, and the best that could be expected was to extract mean values and some
higher-order statistics. Even quite a few years later \cite{frisch91}, it was possible to refer to
turbulence research as ``botany'', in the sense of a collection of specimens with little underlying
theory. That this is no longer true of turbulence (nor of botany) is due to the appearance of new
experimental methods and, at least in my personal experience, to the remarkable increase in the
power of computers, which replaced what used to be stochastic variables with tangible structures. It
is probably not a coincidence that the earliest models involving discrete eddies, like Richardson's
\cite{rich22} or Obukhov's \cite{obu41}, or the earliest examples of deterministic chaos
\cite{lorenz:63}, came from atmospheric scientists, because it is mostly in the atmosphere that we
are likely to face turbulent structures `in person'. Everybody has experienced a wind draft, but few
have felt the much smaller sweeps and ejections of an industrial boundary layer. Direct simulation
extended the atmospheric experience to all kinds of flows, and transformed turbulence theory into a
testable proposition.

Most of the paper up to now has been a survey of this transformation. But we have seen that much
remains unknown, especially regarding multiscale interactions, and it is natural to ask whether new
experimental, theoretical or computational techniques can be exploited to attack these problems. A
detailed review of these techniques is unfeasible within the scope of the present paper, but we can,
at least, reflect on what they may mean for turbulence theory, and on what they could offer.

\subsection{Data-driven science}\la{sec:data}

The best known among the new observational technologies are data-driven methods and the different
flavors of artificial intelligence (AI), and the most striking theoretical development has been
causality analysis. All of them have been made possible by the decreasing cost of digital storage.
Computer speed has been discussed often. In the years from 1960 to 2020, the speed of the fastest
computer increased by a factor of $10^{11}$ and, as we have seen above, this allowed us to
identify turbulent structures that were invisible fifty years ago. It has also produced prodigious
amounts of data. Terabyte data sets and Petabyte data bases are today common, and not exclusive of
simulations; time-resolved tomographic particle-image velocimetry produces comparable amounts
of data. Fortunately, the cost of storage has decreased apace. During the period mentioned above,
the price of a random-access Megabyte has decreased by a factor of $10^{10}$.

We mentioned in the introduction that too many data turn direct simulations into black boxes,
because it is difficult to extract information from them. Both data-driven science and AI are
tools to help us make sense of these data. They have been changing the practice of fluid mechanics
in recent years and, to some extent, that of science in general. Causality analysis has been one of
the results. Although the technical details of how AI works are still in flux and will probably
remain so for some time, it is beginning to be possible to discuss whether data-driven methods are a
new paradigm in turbulence research or just an incremental improvement on what we already had.

What these methods assume is that all the information about a system is contained in the statistics
of its observed behavior, and they provide ways in which these statistics can be organized and
inter- or extrapolated, while requiring relatively little understanding on the part of the user. In a sense,
data-driven schemes are successors to the table-lookup and graph-driven methods of the late
nineteenth century, although larger data sets, faster computers, and better algorithms makes them
much more useful now than they were then.

However, there are several problems. The main one is the assumption that the observational data
contain all the information on the system or, equivalently,  that the attractor of a
dynamical system univocally defines the system. This is not true, as can be seen by enlarging the
dynamical system \r{eq:DS0} with another equation
\beq
\dr y/\dr t =-y,
\la{eq:DSdecay}
\eeq
where $y$ is not in $X$. The attractor is always located at $y=0$, and dominates the statistics of
any sufficiently long observation, but it contains little information on the structure (or even on
the existence) of Eq. \r{eq:DSdecay}. This difficulty is common to all systems with an attracting
manifold and, although the attractor is probably always enough to describe the natural evolution of the system, controlling
it, or even predicting uncommon extreme events, requires a fuller understanding of the equations.

Researchers working in medical or social sciences, for which constitutive equations are unavailable,
distinguish between `behaviors' and `mechanisms', approximately corresponding, respectively, to the
observable evolution of the system and to the underlying equations \cite{marin25}. There is little
doubt that mechanisms predict behaviors, at least in principle, but the converse is not necessarily
true. As our above example shows, many different mechanisms can result in the same behavior and, in
that sense, mechanisms are not necessarily derivable from observations. This is why the emphasis in
hypotheses-driven research is on falsifying, instead of on confirming, them. Reference
\cite{marin25} discusses how it may be possible to study behaviors without necessarily understanding
mechanisms in fields in which equations are unknown. This is the black-box approach of AI, but
\cite{marin25} and the references therein also emphasize that predicting, or controlling, behavior
in previously unobserved conditions requires mechanisms.

\subsection{Causality and the return of chaos}\la{sec:cause}

The other recent development in the analysis of complex systems is causality, mainly adapted from
equation-less disciplines such as sociology and economics \cite{Pearl:09,AngEtal:96}. Causality is
often cited as a bedrock of science, one of whose roles is to predict and manipulate the future
using the knowledge of causes in the past, although one of the lessons of quantum mechanics is that
it is possible to have a well-formed theory without strict causality or determinism.

In fields like fluid mechanics, in which equations are known and formally deterministic, it has been
argued that causes are equivalent to initial conditions \cite{russ:12}, so that the identification
of causes ought to be equivalent to the integration of the time-reversed equations of motion. Unfortunately,
even when the forward integration of a dissipative system, such as a viscous fluid, is well-behaved,
the time-reversed equations are ill-posed, and can only be integrated statistically. In this sense,
even if we know the effects of a given cause, the converse is not true, and we can only give a
statistical description of the possible causes of a given effect.
 
Even more, neither is the direct turbulence problem fully deterministic in an operational sense. We
began this paper by describing the Navier--Stokes equations as a deterministic dynamical system, but
this is only partly true because systems with many degrees of freedom are sensitive to infinitesimal
perturbations. In which sense the numerical integration of such a system converges to a solution of
the underlying equations is unclear, but it is common knowledge in the direct simulation community
that recompiling a code with slightly different compilation flags leads to divergence between the
old and the new solutions in approximately one eddy turnover. A chaotic simulation with errors
either in the initial conditions or in the integration scheme can be understood as the evolution of
the correct equations with perturbed initial conditions \cite{hay:jack:07} or as the integration of
a perturbed set of equations \cite{Gri:Sanz:86}. The two errors are different, and not fully
interchangeable, but both grow on average exponentially with the simulation time
\cite{Teix:Judd:07}. Better numerical schemes, better defined initial conditions, or more
precise arithmetic only result in moderately longer simulation times for a given desired error. The
result is that the system is only deterministic in practice over times of the order of the eddy
turnover, which may be different for different flow scales and for different parts of the flow. On
the other hand, although this is true for individual trajectories, it is empirically found that the
simulation of a sufficiently large ensemble of initial conditions results in essentially correct
statistics, although the reason is not well understood \cite{Corless:25}. If the system is ergodic,
the same is true for a long simulation, but only if it is interpreted as a source of statistics,
rather than as the evolution of a particular flow. An application to a simple two-dimensional case
is \cite{VelaAvi24}.

We mentioned in \S\ref{sec:coherent} that reduced-order models are bound to be stochastic, because
they need to incorporate the effect of the neglected variables. This Voltaire \cite{voltaire}
randomness is easy to understand as an acknowledgment of ignorance, but says more about how we
manipulate the dynamical system than about the system itself. The randomness due to chaos is deeper,
and unavoidable in any practical sense because the initial conditions cannot be precisely known and
uncertainty is continuously injected into the simulation or in the experiment by numerical or
mechanical noise. In this sense, all simulations or experiments of a turbulent flow can only be
interpreted statistically over sufficiently long times.

The coherent structures discussed in the previous sections are intended to tame this statistical
uncertainty, but our finding that the coherence time of individual structures is of the order of a
few turnovers implies that this approximate determinism only extends to short times. This does not
mean that causality analysis is impossible, but it suggests that the equations of motion should be
interpreted as evolution equations for probability distributions in state space. Such a
representation was first proposed by \citet{hopf52}. Individual flow states are replaced by the
probability distribution of ensembles of flows, and the equations of motion become a linear and
reversible evolution operator that acts on the distributions. Unfortunately the dimensionality of
the space of distributions is much higher than for the space of states. For example, if we consider
a single variable $u$ and its distribution, $p(u)$, computed over $m$ bins, the dimensionality of
$u$ is $D_u=1$, while that of $p(u)$ is $D_p=m$. This made this type of statistical analysis
impractical for a long time, but recent improvements in computer power are starting to make it
possible for very-reduced-order models \cite{jimPF23,jimJFM24}.

There are two main definitions of causality, both of which are easily adapted to a probabilistic
setting. In the strict sense, $A$ is part of the cause of $B$ if modifying $A$ also modifies $B$
\cite{Pearl:09}. Determining strict causality is necessarily interventional, because the experiment
of modifying $A$ actually has to be carried out. In some cases, this may be impossible, as in
forensic investigations in which the goal is to find the probable cause of a past event. In other
cases, it may simply be very expensive, because many possible modifications have to be tested on
many putative events. In addition, one can never be sure that all the relevant experiments have been
performed, so that, while it may be possible to prove that $A$ is causally important for B, proving
that it is irrelevant may be harder. Even so, the growth of modern computational power is starting
to make such Monte-Carlo searches practical in actual flows, leading in some cases to unexpected and
interesting causal relations \cite{jimenez21,osawa24}.

We have referred several times in our discussion to a variant of this interventional approach in
which the equations of motions are modified to inhibit the influence of whole classes of putative
structures \cite{jim:pin:99,tue:jim:13,loz:etal:21}. These conceptual experiments are useful because
they address dynamical paths, rather than individual structures, but they share with strict
causality the high cost of doing experiments, and they are not always easy to interpret because a
process usually affects many different parts of the flow. On the other hand, it can be argued that
dynamics, rather than structures, is the most interesting part of causality.

An alternative to alleviate the cost of Monte-Carlo causality searches are causal-empirical methods
\cite{AngEtal:96}, in which experiments are substituted by statistics conditioned to some particular
combination of events. The conditioning event is treated as a natural experiment, and the evolution
of the actual system is substituted by a data-driven model. These methods were pioneered by
\citet{granger:69} in economics, and have become very popular because of their savings with respect
to strict causality, but an analysis of the possible pitfalls is given in \cite{Pearl:09}, and they
share with all data-driven methods the caveats raised in our previous section about the relation
between a system and its attractor. An application to fluid mechanics that addresses some of these
drawbacks is \cite{SanArrLoz:24}, and it is interesting that their results regarding inner-outer
interactions in wall-bounded turbulence are compatible with those of the interventional experiments
in \cite{osawa24}, in spite of the very different methodologies.

\section{Conclusions}\la{sec:conc}

We have seen that our understanding of turbulence has been revolutionized in recent years, in the
sense that many statistics have been replaced by detailed dynamics. I have given examples for two
particular kinds of shear flows, and shown that the resulting rationalizations are not only
intellectually attractive but technologically useful, especially in the area of control. However, we
have also seen that many problems remain open, particularly regarding multiscale turbulence (which,
of course, is the only kind). Actually, many of the mechanisms described here apply to simplified
situations, such as minimal flow units, and many of the open questions appear when we try to
bring them closer to real multiscale flows.

Turbulence has always had to balance science with applications, and physics with engineering and,
although not originally planned that way, this review has tilted towards physics. This may be
because this is where more progress has taken place lately, but anybody reading this review may
also conclude that beauty has been given more weight than applications. I plead guilty of this bias,
which already appears in the personal list of reasons to keep studying turbulence included in the
introduction. I defend it. Historically, many scientific theories that started as intellectually
attractive curiosities have resulted in technological breakthroughs, although often in areas
different from the ones they were intended for. Others have not. It is a fundamental property of
interesting natural laws that they apply to situations beyond those for which they are developed
\cite{Feynman65} and, unfortunately, it is usually impossible to determine in advance which
beautiful theories will also turn out to be interesting.

I have argued that the theoretical transformations of the past few years have been made possible by
new experimental and computational methods, in my case especially the latter, and, in surveying
which new techniques might be useful for the problems that remain open, I have centered on
data-driven science and in the search for causality. Both have proved to be powerful enough in other
scientific disciplines to expect them to be useful in turbulence and, although they are probably not
the only ones that should have been considered, they are all that can be addressed in the limited
space of this review.
But it is important to understand their limitations as well as their strengths. Artificial
intelligence is an answering machine, and answering questions is not the hardest, nor even the main
problem in research. Much harder is usually to identify interesting questions. Some uses of AI
may actually hurt research because they discourage new questions by providing quick answers.
Data-driven methods typically work by minimizing some error measure between predictions and data,
but the optimum is seldom unique, and the answer is usually located in state space near the question
being posed. Asking a better question is essentially providing a new initial guess that offers the
chance of finding a better optimum. It is not clear how this is currently done by successful
researchers, nor how can it be automated, although some of the computer gaming successes of AI
suggest that it might be possible \cite{Alphago16,solera:etal:24}.

Some AI practitioners do not seem to care about all this, or about questions beyond the empirical
behavior near the system attractor (although generative AI claims, and appears, to do more than
that). This is troubling at first sight, but these users may have a point. The worries about black
boxes recall older arguments about new ways of doing things. A relatively recent example is the
discussion about whether computer-assisted proofs in mathematics are proofs at all
\cite{fraser:etal:24}, or whether numerical simulations in physics should be considered more than
examples waiting for an analytic solution, but the argument is much older. More than twenty
centuries ago, Plato \cite{PlatoPhaedrus2005} complained about written arguments, as opposed to live
dialogues, because ``\ldots\ books appear to talk to you as if they were intelligent, but if you
want to know more, and ask them something, they repeat the same thing over and over again''. The
current arguments about the limitations of artificial intelligence have a similar flavor, and may
suffer a similar fate.

Table lookup has been a way of doing engineering for a long time, and was still the preferred way
fifty years ago. For fluid mechanics, we saw in the introduction that the equations of motion were
written in the flowering of rationalization during the Enlightenment, but could not be used
effectively because computers were inadequate or nonexistent. The rise of high-speed computers
during the second world war brought the equations closer to applications, eventually resulting in
direct and large-eddy simulations, but it is debatable whether understanding structures and
dynamical systems should be part of the same computation as predicting drag. We argued in
\S\ref{sec:data} that AI is a throw-back to table lookup with new technology, and it may be the case
that the use of the constitutive equations for routine engineering applications has been an
interlude between tables, with little more than empirical support, and the newer AI systems that
summarize full computations.

It is possible to envisage a state of affairs in which most engineering calculations are outsourced
to data-driven AI systems, while theory, direct and even large-eddy simulations, possibly aided by
AI techniques, maintain their role of enlarging the frontiers of fluid mechanics and of developing
novel engineering and control solutions. In essence, AI would deal with behaviors, and equations
would take care of mechanisms. Less clear is whether generative AI might one day be able to take
over some of the discovery activities. I was trained as an applied mathematician and, if that were
the case, I would prefer the machine to communicate with us in the language of equations. But this
is a personal preference that may not be optimal from the AI's point of view, and it is conceivable
that we may have to learn new ways to communicate, or make do with translated text. Think of the 
difficulty of explaining Bourbaki mathematics \cite{Bourbaki06} in the notation of  Isaac Newton.\vspace{-1ex}

\begin{acknowledgments}
This work was supported by the European Research Council under the Caust grant ERC-AdG-101018287. I
am grateful for discussions with many colleagues, particularly Marc Avila, Jos\'e I. Cardesa, Peng
Chen, Ana Crespo, Miguel P. Encinar, Ricardo Garc\'ia-Mayoral, Petros Ioannou, Adrian
Lozano-Dur\'an, Yvan Maciel, Vera Pancaldi, Juan M. Rojo, Jes\'us M. Sanz-Serna, Julio Soria and
Ricardo Vinuesa.
\end{acknowledgments}

%

\end{document}